# *On the role of the plasmodial cytoskeleton in facilitating intelligent behaviour in slime mould Physarum polycephalum.*

Richard Mayne, Andrew Adamatzky and Jeff Jones

International Centre of Unconventional Computing

University of the West of England, Bristol, UK

Tel.: +447979 522256

{richard.mayne}/{andrew.adamatzky}/{ jeff.jones }@uwe.ac.uk

## *Conflict of Interest Statement*

The authors declare no conflict of interest.

## *Abstract*

The plasmodium of slime mould *Physarum polycephalum* behaves as an amorphous reaction-diffusion computing substrate and is capable of apparently 'intelligent' behaviour. But how does intelligence emerge in an acellular organism? Through a range of laboratory experiments, we visualise the plasmodial cytoskeleton – a ubiquitous cellular protein scaffold whose functions are manifold and essential to life – and discuss its putative role as a network for transducing, transmitting and structuring data streams within the plasmodium. Through a range of computer modelling techniques, we demonstrate how emergent behaviour, and hence computational intelligence, may occur in cytoskeletal communications networks. Specifically, we model the topology of both the actin and tubulin cytoskeletal networks and discuss how computation may occur therein. Furthermore, we present bespoke cellular automata and particle swarm models for the computational process within the cytoskeleton and observe the incidence of emergent patterns in both. Our work grants unique insight into the origins of natural intelligence; the results presented here are therefore readily transferable to the fields of natural computation, cell biology and biomedical science. We conclude by discussing how our results may alter our biological, computational and philosophical understanding of intelligence and consciousness.

# Introduction

Slime mould *Physarum polycephalum's* vegetative life cycle stage, the plasmodium (plural plasmodia, Fig. 1), is a macroscopic, multinucleate, acellular organism which behaves as a living amorphous reaction-diffusion computing substrate.[1] To delineate, slime mould may be considered as an unconventional computing substrate in which data are represented as transitions in chemical equilibria in an excitable medium. The plasmodium is able to concurrently sense input from a range of stimuli including temperature, light, chemicals, moisture, pH and mechanical force. [2–5] *P. polycephalum's* innate behaviour patterns may be manipulated experimentally to perform useful computational tasks, such as to calculating the shortest route between any number of spatially distributed nutrient sources, or navigating its way out of a maze via chemotaxis; these operations may be interpreted in terms of computational geometry, logic and spatial memory. [6,7]

We are rapidly approaching the physical limitations of the materials used in the creation of traditional solid state computers, and their manufacture is ecologically-damaging. For these reasons, research into unconventional computing is gathering momentum. Unconventional, or non-classical computation, (UC) utilises the natural properties and processes of physical or living materials to provide useful computational functions. These systems are typically composed of simple and plentiful components, contain redundant parts (i.e. not being dependent on highly complex units), and show resilient or 'fault tolerant' behaviour. UC is often observed in systems which show 'emergent behaviour', novel behaviour which emerges from the interactions between simple component parts and which – critically – cannot be described in terms of the lower level component interactions. Emergent behaviour is found in systems with many simple, local interactions and which display self-organisation, i.e. the spontaneous appearance of complexity or order from low-level interactions. Many of the attractive features of UC computing devices (distributed control, redundancy, fault tolerance) are generated by mechanisms of self-organisation and emergence, and the study of these properties is useful not only from a computational perspective, but also from a biological viewpoint – since much of the complexity in living systems appears to be built upon these principles.

UC is attractive because, for a number of applications at least, utilising the natural properties of physical systems for computation is a much more efficient means of computation. Non-classical computation can take advantage of parallel propagation of information through a medium (for example in the chemical approximation of Voronoi diagrams, or the parallel exploration of potential path choices in path problems using microfluidic plasmas). [8,9] Slime mould is an ideal living candidate for a UC computing medium because it can behave as a distributed computing 'material' capable of complex sensory integration, movement and morphological adaptation, but one which is also composed of relatively simple components that are amenable to understanding, external influence and control.

For some applications UC devices are far superior to solid state architectures in terms of computational power, error resistance and energy efficiency. Furthermore, the UC sub-field of biological computing has great relevance to human medicine due to the essentially computational nature of the brain's functional units (neurons). Indeed, many analogies may be drawn between neurons and slime mould plasmodia, such as electrical excitability, branching/stellate morphology and memristivity (the basis of synaptic plasticity, and thus memory).[10] Neuron culture methods are expensive, temperamental and ethically restrictive, however; slime mould culture techniques are comparatively simple, cheap and resilient, with no associated ethical issues.[11]

Lacking any biological components normally associated with so-called 'intelligent behaviour' (brain, neurons etc.), the physiological origin of *P. polycephalum's* computational abilities are still largely unexplained, and as such, we are denied a vital tool for manipulating slime mould behaviour to our advantage. Several authors have previously formalised slime mould behaviour patterns with mathematical models of protoplasmic network dynamics, which, whilst valuable in their description of how protoplasmic computations are enacted out at the organismal level, tend to adopt a top-down view of the phenomena observed and do not explain the underlying basis of the computation involved. [12–14] This investigation was designed to explore the basis of slime mould intelligence from a bottom-up perspective, at the molecular level.

We begin by proposing that the *P. polycephalum* plasmodium must possess some form of network through which sensory and motor data may travel, as information transmission is a fundamental component of every system capable of computation (partnered with the means for data storage and logical processing of data). Furthermore, without such a network, incoming data – defined here loosely as an environmental stimulus which elicits a response within the organism – would be 'unstructured', data structuring here meaning the production of predictable, quantifiable data patterns from unstructured environmental data.

Indeed, in Ref. 15, Lungarella and Sporns argue that the dynamical coupling of sensorimotor data streams with the morphology (here meaning both structure and physical properties) of a computational entity's (biological or artificial) data network induces automatic information structuring, which in turn allows the entity to dynamically react to its environment: this is thought to be the underlying basis of learning and logical ability. [16–19]

Consequently, as the morphology of these data streams define how the entity may sense and interact with its environment, it essentially carries out a proportion of the computation. As such, computational processes are 'outsourced' to the morphology automatically, which reduces the workload of the entity's control unit/s. [15,16,20–22]

This view of natural computing is derived from the precepts of 'morphological computation', a concept usually employed in robot design, wherein entity compliance is exploited to create self-stabilising systems which reduce the need for constant monitoring by the control unit.[16,22] By amalgamating biological and computer sciences, recent advances in the field have shown promise in modelling natural systems which have not been fully described in purely mathematical (algorithmic) terms, e.g. brain function.[23]

Whilst assigning a tangible structure to the abstract term 'data stream' may appear to be a simplification for the ease of description, it is clear that some form of network for optimising intracellular communication is present within slime mould, as the phenomena we may label as 'inputs' (i.e. environmental sensing) display redundancy, rapidity and transduction into different formats. For example, cellular signalling events may be observed to propagate far faster than simple diffusion of signal molecules through the cytoplasm would allow for, and many stimuli are transduced into multiple different formats, e.g. mechanical stimulation of a cell can provoke the generation of biochemical and bioelectrical signalling events.[24]

Slime moulds would also appear to have a unique necessity for a data network as a single plasmodium can contain many millions of nuclei. Whilst likening the functions of a cell's nucleus with those of a processor is a contentious issue (which is not discussed here), it is clear that the nuclear processes which alter the cell/the cell's behaviour (activation/repression of genes, induction of signalling cascades etc.) are direct consequences of the environmental data it receives. How, then, are the activities of millions of nuclei are synchronised to produce coherent behaviour? The existence of a plasmodial data network would appear to be an elegant solution to this problem, if such a network were demonstrated to connect nuclei together (hence facilitating internuclear communications) and/or interconnectivity (to allow for signal amplification).

All eukaryotic (and some prokaryotic) organisms possess an intracellular network; a plentiful protein scaffold known as the cytoskeleton, which is composed of tubulin microtubules (MTs), actin microfilaments (MFs) and a range of intermediate filaments (IFs) (Fig. 2). Tubulin and actin are ubiquitous, whilst IF type differs depending on the function of the cell. All three cytoskeletal protein groups are considered to interlink with each other and with most of the major cellular organelles and receptors, to form one single interconnected network.[25] The cytoskeleton participates a multitude of cellular functions, many of which are considered to be essential to life; these include mechanical rigidity, motility and substance (organelle and molecular) trafficking.[26]

It is becoming increasingly evident that the cytoskeleton also participates in a wide range of cellular signalling events. The forms of information it may transmit include electrical potential, mechanical force, quantum events, propagating waves of protein conformational changes, facilitated chemical transport and biochemical signal transduction cascades. [24,26–32] The cytoskeleton therefore represents an attractive model for describing data transmission within cells, as all of these signalling events may be interpreted as data being fed into a cellular computer. We are by no means the first to suggest this, but such a topic has not been previously described in slime mould models.

This study was undertaken to explore the putative role of the slime mould cytoskeleton in facilitating intelligent behaviour by acting as an intraplasmodial data network. This was achieved by visualising the plasmodial cytoskeleton with confocal microscopy and basing a range of computer modelling techniques on the structural observations made.

## *Results*

### *Visualising and formalising the plasmodial cytoskeleton*

The actin and tubulin components of the *P. polycephalum* cytoskeleton were visualised with confocal microscopy, see Figs. 3 and 4 (see Methodology section for details of sample preparation). Samples were taken from the two distinct anatomical locations of the plasmodium, namely the tubular plasmodial 'veins' that form the posterior areas of the plasmodium and the 'fan-shaped' advancing anterior margin formed from converging pseudopodia (see Fig. 1). The actin and tubulin components of the plasmodial cytoskeleton appear to be extremely complex, highly interconnected networks – especially so when compared to animal cell counterparts. The actin MF network is extremely abundant in the advancing pseudopodia (likely to result from the propulsion of motile machinery to the pseudopodial growth cone to facilitate movement), although somewhat less so in plasmodial veins (data not shown); the tubulin network would appear to be profuse in both anatomical locations, but less so than actin in the anterior portions of the organism. Both cytoskeletal proteins appear to articulate onto each nucleus at several points and are also present at the external membrane, indicating associations with surface receptors (as has been demonstrated in animal cells).[26]

Both tubulin- and actin-component cytoskeletal networks can be formalised using proximity graphs if edges are modelled as the cytoskeletal network linking the nuclei, nodes. Note that nuclei are here likened as nodes to reflect their putative computational function, not as major structural elements of the cytoskeleton. A proximity graph is formed under the principle that two nodes are connected by an edge if they are spatially close: each pair of points is assigned a certain neighbourhood, and points of the pair are connected by an edge if their neighbourhood is empty. Exemplar graphs of two families – Gabriel graph and relative neighbourhood graph – are shown in Fig. 5ab. Nodes $a$ and $b$ are connected by an edge in the relative neighbourhood graph if no other point $c$ is closer to $a$ and $b$ than $dist(a,b)$.[33]

Nodes $a$ and $b$ are connected by an edge in the Gabriel Graph if disc with diameter $dist(a,b)$ centred in middle of the segment $ab$ is empty.[34,35] As proved by Toussaint, the so-called Toussaint hierarchy, the minimum spanning tree – an acyclic graph spanning all planar data nodes with minimal sum of edges – is a subgraph of the relative neighbourhood graph, which is a subgraph of the Gabriel graph (Fig. 5c); all are subgraphs of the Delaunay triangulation.[33] We speculate that any particular graph in the Toussaint hierarchy might be employed in intraplasmodial computing, to meet the immediate needs of the plasmodium. For example, the minimum spanning tree might be used for energy-economic broadcasting and implementation of basic decision-making tasks, e.g. inference of data. The Gabriel graph could be used for majority voting when comparing data obtained from different receptor sites, as such graph topologies have been demonstrated to complete this type of computation efficiently.[36] The co-existence of different graph topologies within a single plasmodium may imply that it is able to switch network topology to meet the physiological demands of the organism.

Whilst proximity graphs would appear to be good models for cytoskeletal network topology, it would also appear that, in the absence of any clearly defined rules for network structure, the combined actin and tubulin network may also be described in terms of a random graph at the anatomical locations where the network is densest. The random graph is a useful approximation for systems which are not fully understood as their network dynamics are computed probabilistically (which is particularly relevant here, as cytoskeletal assembly is thought to be a stochastic process), although it has been noted that unmodified random graphs, as per the descriptions of Erdős and Rényi, are not ideal models for real-world networks; for example, no allowances are made for phenomena such as clustering (transistivity). [37–39] Modifications to the model exist, however, such as the inclusion of a clustering coefficient by Watts and Strogatz in Ref. 40, and indeed node clusters/subgraphs would appear to be abundant towards the peripheral regions of the plasmodium.[41]

This model also allows for signal percolation through the cytoskeleton, which would simultaneously allow for the conservation of cost (in this case, energy) by reducing the number of necessary network nodes (nuclei) whilst preserving maximum network functionality in the event of nuclear rearrangement --- a process which occurs constantly in the plasmodium during migration.[24,42] Consequently, emergent behaviour may be observed in artificial electrical random networks of mixed resistive and reactive entities, thereby providing another basis to explain the occurrence of emergent behaviour within the plasmodium.[43]

### *Outsourcing computation to the cytoskeleton*

Let us apply paradigms of morphological computation (see Introduction) to a hypothetical slime mould under the assumption that the cytoskeleton functions as a data network. Stimulation of surface receptors, e.g. by detection of a chemical gradient, may prompt the signal to be transduced as electrical potential/quantum events/molecule transport (or any combination of these) through the cytoskeletal proteins coupled to the receptor. The frequency (and also possibly the magnitude) of incident signals are proportional to the degree of stimulation, but are transduced in an unambiguous, statistically repeatable format, and are then transmitted to the parts of the organism which receive and act upon such information (here assumed to be the nuclei). It is assumed that when the hypothetical signal reaches a branch in the cytoskeletal network, the signal may propagate down multiple branches, depending on its 'pattern' (encompassing data type, frequency and possibly magnitude), thereby amplifying the signal to inform multiple nuclei. Hence, data has been converted into a format the slime mould can 'understand' and broadcast to multiple nuclei, despite the process requiring no control from the slime mould: these tasks can therefore be said to have been outsourced to the morphology, in that they occurred automatically as a consequence of the physical properties of the cytoskeleton.

The slime mould has no need to continuously sample receptors for data, which simply generate their signals automatically when required; the structuring and format of the signals are also automatic, removing the need for slime mould controllers to manually generate and control the physics of its sensory system. When compared with classical computing architectures whose programs need to manually define the way in which it interacts with its environment, the slime mould model would appear to be extremely efficient. Furthermore, if the stimulus does indeed propagate to multiple controllers, parallelism is achieved at no extra energy cost. There is also no need for the slime mould to control nucleus synchronisation if the network topology favours majority voting, as previously mentioned. The resulting system saves energy yet enhances the slime mould's computational potential; this makes perfect evolutionary sense and hence enhances the feasibility of the model.

As the slime mould moves forwards, its sensory apparatus and cytoskeleton are re-shaped automatically to fit the new form of the organism. This highlights how the structure of plasmodial data streams are dynamically coupled to its environment. To an extent, the plasmodium also reciprocally interacts with its environment's morphology, e.g. by following nutrient gradients from food sources before altering them as the food is consumed.

These hypotheses provide a theoretical basis to describe how slime mould may be able to behave in an apparently intelligent way (i.e. carry out complex computations) whilst lacking any well-defined apparatus usually associated with such phenomena, i.e. a brain or nervous system.

### *Modelling information transmission through the cytoskeleton*
**Automaton modelling**

By analysing the structure of actin filaments (Fig. 6), we may begin to model hypothetical information transmission events propagating through MFs with two-dimensional cellular automata. Variations of this method have been used previously to model tubulin MT information transmission.[44,45] This section details a preliminary model to describe how computations may occur if cytoskeletal 'data' are considered to be hypothetical generalised energetic events which propagate from one molecule to the next according to the following rules.

The actin automaton consists of two chains $x$ and $y$ of nodes (finite state machines), as shown in Fig. 6. Each node takes three states: excited (+), refractory (−) and resting (∘). Practically, these states are modelled to correspond with the quantum mechanical events which occur when a molecule undergoes energy state transitions: 'excited' corresponds with a raise in energy level as the signalling event is transmitted to the actin molecule, and refractory corresponds with a decrease in energy level as the signal is transmitted to adjacent molecules. Resting refers to the ground state of the molecule when it is unstimulated. Note that this model is generalised to apply to a wide range potential signalling events, e.g. chemical energy conferred from ATP hydrolysis, electron transport across each molecule.

A node updates its states depending on states of two closest nodes in its chain and the two closest nodes in the complementary chain. Thus neighbourhood $u(x_i)$ of node $x_i$ is a tuple $u(x_i) = (x_{i-1}, x_{i+1}, y_{i-1}, y_i)$ and neighbourhood $u(y_i)$ of node $y_i$ is a tuple $u(y_i) = (y_{i-1}, y_{i+1}, x_i, x_{i+1})$ (Fig. 6). Let $z$ be either $x$ or $y$ and $\sigma(u(z^t)) = \sum_{p \in u(z)} \{p^t = +\}$ is a number of excited neighbours of node $z$ at time step $t$. All nodes update their states by the same rule and in discrete time:

$$z^{t+1} = \begin{cases} +, & \text{if } z^t = \circ \text{ and } C(\sigma(u(z_i)^t)) \\ -, & \text{if } z^t = + \\ \circ, & \text{otherwise} \end{cases}$$

where $C(\sigma(u(z_i)^t))$ is a predicate on a the neighbourhood state. A resting node becomes excited if it has a certain number of excited neighbours, an excited node takes refractory state and a refractory node takes resting state. The following predicates are considered:

1. $C_1(\sigma(u(z_i)^t)) = <\sigma(u(z_i)^t) > 0: z = x, y>$: a resting cell in each chain is excited if at has at least one excited neighbour;

2. $C_2(\sigma(u(z_i)^t)) = <\sigma(u(z_i)^t) = 1: z = x, y>$: a resting cell in each chain is excited if at has exactly one excited neighbour;

3. $C_3(\sigma(u(z_i)^t)) = <\sigma(u(x_i)^t) > 0 \text{ and } \sigma(u(y_i)^t) = 1>$: a resting cell in chain $x$ is excited if at has at least one excited neighbour and a resting cell in chain $y$ is excited if at has exactly one excited neighbour.

Chains evolving by $C_1()$ show classical dynamics of one-dimensional excitation waves (Fig. 7). In the space-time configuration illustrated in Fig. 7, there are originally four sources of excitation waves, from which eight waves develop. Most of them collide with each other and extinguish and two waves travel towards boundaries of the chain.

Rule $C_2()$ is analogous to $2^+$-rule developed in Refs. 46–48 because nodes react only on a specific number of excited neighbours. The space-time excitation dynamics of rule $C_2()$ is very rich (Fig.8). It exhibits a range of elementary and composite travelling excitations, analogs of gliders in cellular automata, and generators of the mobile localisations, analogs of glider guns. As clearly shown in (Fig. 8), the following scenarios of collisions can be observed: composite travelling excitations are cancelled by standing excitation wave, colliding travelling localisations produce localisation generators, one of the colliding localizations is reflected or cancelled in the result of collision while other localisation continues travelling undisturbed. This range of collision outcomes is enough to implement collision-based logical circuits.[49]

Model $C_3()$ assumes two chains of actin molecules follow different rules of excitation. This may be questionable from a physical perspective, but is worse to consider in computer models: random excitation of such chains leads to the formation of stationary generators of travelling excitation waves in both chains (Fig.9). Frequency of generation is higher in chain $x$. Chain $y$ exhibit rare travelling defects — disturbances of underlying or background excitation patterns — as one travelling east in Fig. 9b. These defects can be also used to implement collision-based circuits in actin automata.[49]

Crucially, the patterns observed Figs. 7-9 arguably show emergence, that is, (subjectively) complex patterns emerge from simple input patterns, as well as the basis for implementing basic logical circuitry. This provides a theoretical basis to describe how complex behaviours my result from cytoskeletal communications which participate in plasmodial computation. It is prudent to note that this model does not account for the actin network branching, and that MFs articulate onto each other via actin-binding proteins (see Fig. 2), through which the physics of data transmission may differ, potentially creating another level of complexity.

Further work will therefore focus on encompassing these issues into the model, as well as modelling information transmission through tubulin MTs and IFs.

# Discussion

### *Morphological Adaptation and Outsourced Computation in a Model of Slime Mould*

We introduced an approach to modelling emergent morphological computation in slime mould in Ref. 50, which consisted of a large population of simple components, mobile agents, with simple local interactions. The agent particles were placed on a discrete lattice and were coupled indirectly by overlapping sensors, the particles represented random arrangements of cytoskeletal filaments. Agents sensed the concentration of a hypothetical 'chemical' in the lattice, oriented themselves towards the locally strongest source and deposited the same chemical during forward movement. The initially random arrangement of particles coalesced into emergent transport networks which underwent complex evolution, exhibiting minimisation and cohesion effects under a range of sensory parameter settings. Nutrients were represented by projecting values into the lattice at fixed sites and the network evolution was affected by nutrient distribution and nutrient concentration. The collective behaved as a virtual material whose morphological adaptation to its environment reproduced the behaviour and spatially represented computation seen in slime mould. The model has been shown to spatially compute proximity graphs, Voronoi diagrams, combinatorial optimisation, collective transport, and collective amoeboid movement.[50–54]

Critically the agent model uses the same principles, and apparent limitations, of slime mould; simple component parts and local interactions. The model utilises self-organisation to yield emergent behaviour and an embodied form of (virtual) material computation as seen in slime mould. The embodied computation seen in the model also employs the environment as a means of outsourcing computation. In fact the importance of environmental stimuli cannot be overemphasised in this modelling. Without any environmental stimuli the model simply generates reaction-diffusion patterning. It is the stimuli provided by external attractants (or repellents) which force the material to adapt its spatial behaviour. The specific mechanism utilised is the diffusion of attractants (or repellents) within the environment. The presence of these stimuli at the periphery of the material provides the impetus for its morphological adaptation. The interaction between environment and the material is two-way, however: When the model population engulfs a nutrient source, the diffusion of nutrients from that source is suppressed. This changes the local configuration of chemoattractant gradients in the lattice (as demonstrated by the changing concentration gradient profiles in Fig. 10e-h as a spanning tree is constructed) which ultimately changes the spatial pattern of stimuli offered to the model population. This mechanism is an efficient use of the environment as a spatial storage medium of stigmergic cues and `offloads' some complex computation to the environment. This exploitation of environmental computation may partially explain how slime mould, with its limited computational 'components' can perform such complex behaviours without a complex nervous system or indeed any specialised neural tissue.

Note that the granularity of the agent model is limited to the scale of individual units of gel/sol interaction, represented by the particles' bulk approximation of filament orientation, aggregation and transport. The model does not (and cannot) represent molecular or quantum interactions in the cytoskeleton and thus may lack some of the speed of information propagation within the organism that such mechanisms would exploit. The speed of computation by the model is limited by the speed morphological adaptation by material relaxation phenomena and a future representation of faster putative cytoskeletal signalling may extend the computational power of the model.

### *On the nature of slime mould intelligence*

Our model of slime mould intelligence has interesting knock-on implications for our understanding of animal intelligence. Other authors have suggested that the metazoan neural cytoskeleton contributes to the computational process – e.g. as a basis for dynamical coupling to the environment via potentiation, or as a physical conductor of electricity – but it is generally considered that intelligent features (memory, cognition etc.) arise from the structure of neural networks and the characteristics of the junctions between them (synapses). [26,55]

We find it feasible to hypothesise that the plasmodial cytoskeleton may have similar functions to those of the mammalian neural network, and by extension, that emergent (and therefore intelligent) behaviour may arise from generalised communications networks, as opposed to exclusively neural tissue. In a sense, it could be considered that the plasmodial cytoskeletal communications network has convergently evolved to similar function.

This poses the research question "is a slime mould as/more intelligent than an animal of a similar size, if we assume that intelligence is proportional to data stream network (neural or cytoskeletal) complexity, abundance and the efficiency of communications therein?". For example, imagine a minute animal such as the Barbados threadsnake (*Leptotyphlops carlae*) whose entire body occupies a volume similar to that of an American quarter dollar coin. A healthy plasmodium may occupy a greater space than such an animal, and furthermore the plasmodial cytoskeleton will be physically larger and significantly more interconnected, as it is not confined to individual cells. Whilst it is true that brain size is not necessarily correlated with intelligence, myriad factors such as the efficiency of communication between components and the informational capacity of the network are likely to be significant factors. Such studies into comparative protist/animal intelligence would surely be of great value, as were slime mould proven to be superior to animals at performing certain tasks, this would challenge some of the existing paradigms for defining 'intelligence', which is essentially a subjective term.

Under the model presented here, slime mould is little more than an automaton: it may only respond to a stimulus in one predefined way, as determined by the physical properties of its components and the state of its environment. A small snake, however, might (perhaps mistakenly) be attributed with somewhat more control over its actions and senses than a slime mould by virtue of being subjectively more 'complex', as a multicellular organism. If slime mould computational abilities were proven to be comparable to those of such an animal, and hence that slime mould cytoskeletal networks are a viable alternative to the neural network for facilitating intelligence, the model presented here would imply that all life is essentially physically deterministic, i.e. an organism can only ever react in one predetermined way to a stimulus.

Finally, if the plasmodial cytoskeleton is considered as an organ enabling intelligence, it represents a casual refutation of traditional Cartesian dualism, or the 'mind-body' problem, as we have presented slime mould as an entity with no distinct separation between its (albeit simple) mind and body. Indeed, we have not presented slime mould with a mind at all, although it is clearly capable of logic, decision making, memory etc. If plasmodial intelligence is demonstrated to be comparable to animal intelligence, are we to imagine slime mould to be possessed of a mind, and by extension, consciousness? And if so, what does that imply regarding our own consciousness?

## *Methodology*

### *Tissue preparation*

Experimental plasmodia were cultured upon 2% non-nutrient agar in the dark at room temperature, and were fed with porridge oats. Samples were prepared via fixation in a solution of 2% Paraformaldehyde and 0.1% glutaraldehyde in pH 7.4 potassium phosphate buffer, dehydration in graduated concentrations of ethanol and clearing in limonene, before embedding in paraffin wax. Blocks were sectioned with a Leica RM2235 rotary microtome at 12μm. All sections were mounted on APES (3-aminopropyltriethoxysilane) coated slides.

### *Immunofluorescence & confocal microscopy*

Slides were de-waxed and underwent antigen retrieval in pH 6.0 0.1M sodium citrate buffer for 30 minutes. They were then cooled in a Coplin jar containing slightly warmed pH 7.4 phosphate buffered saline (PBS). This was followed by exposure to the primary antibody for one hour, three thorough rinses in PBS, exposure to the secondary antibody for an hour and a final series of rinses. Samples were mounted in antifade reagent containing the fluorescent nucleic acid stain DAPI (4',6-Diamidino-2-phenylindole dilacetate) (Molecular Probes, USA) and imaged with a Perkin-Elmer UltraView FRET-H spinning disk confocal microscope.

The primary antibodies used were KMX-1 (anti-myxamoebal β-tubulin; Millipore, UK) at a concentration of 4μg/ml and ACTN05(C4) (anti-actin; Abcam, UK) at 2μg/ml. Secondary anti-IgG antibodies conjugated to Alexa Fluor 488/568 (Life Technologies, USA) were used at a concentration of 2μg/ml.

# *Appendices*

## *Acknowledgements*

The authors would like to thank Dr. David Patton, Mr. David Corry and Mr. Paul Kendrick, all of the University of the West of England, for their technical support throughout this program of research. Mr. Mayne also extends his thanks to Dr. Ben De Lacy Costello for his contributions towards the project's supervision.

## *Photomicrography and Image Processing*

Confocal micrographs were produced with a Perkin Elmer UltraView ERS FRET-H spinning disk laser confocal microscope, and were post-processed with Volocity (colour assignment, noise removal, deconvolution, and brightness and colour adjustments). Unprocessed image files will be made available on request.

# *Figure Legends*

### *Figure 1*

Plasmodium of slime mould *P. polycephalum* growing on an agar-filled Petri dish, feeding on porridge oats. Note the differences in morphology between medial/posterior plasmodial 'veins' (black arrow) and the 'fan-shaped' anterior margin (white arrow).

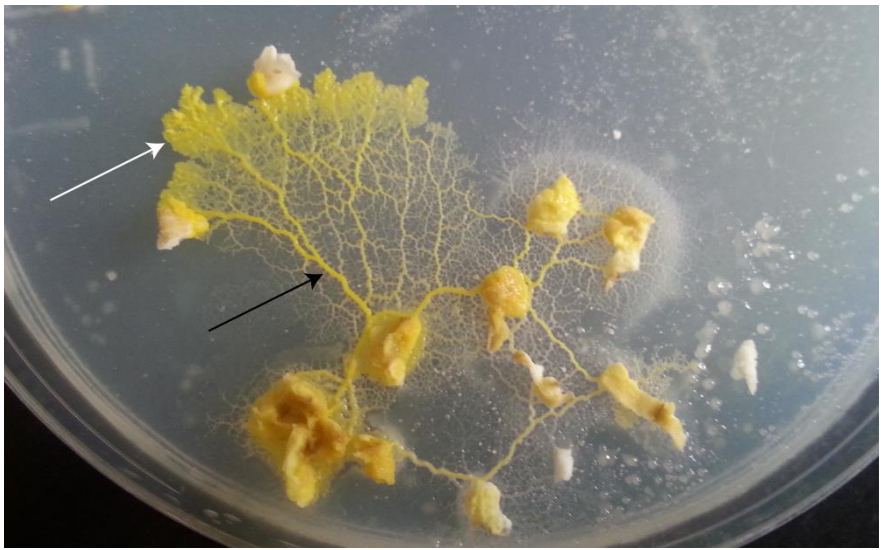

*Figure 2*

Simplified schematic representation of the cellular cytoskeleton in a generalised eukaryotic cell, illustrating how several varieties of cytoskeletal protein form an interconnected network which links cellular organelles together. Intermediate filaments and centrosome are not shown. Nu: nucleus; OM: outer membrane; MAP: microtubule-associated protein; MT: microtubule; ABP: actin-binding protein.

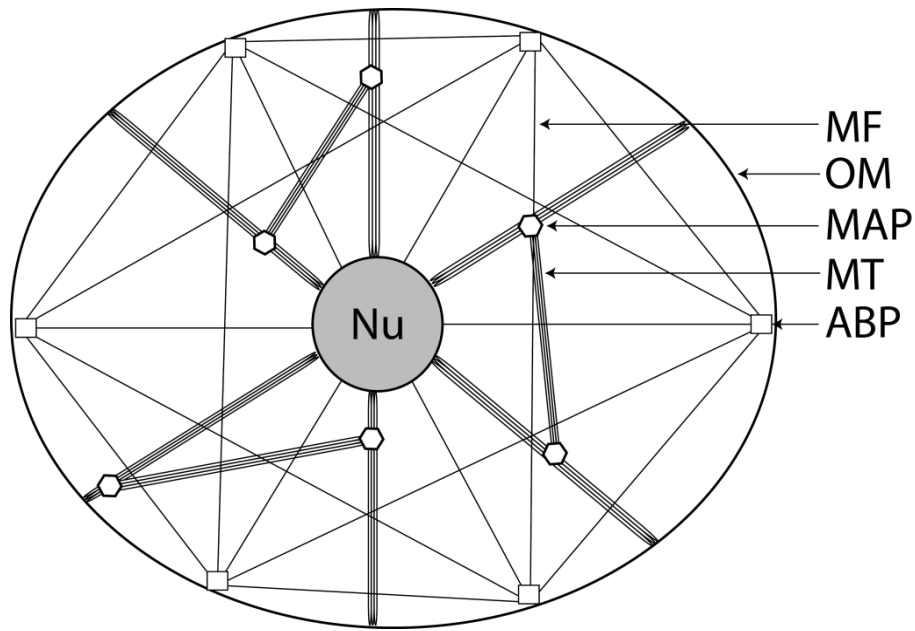

*Figure 3*

Low magnification confocal micrographs of 12μm sections of plasmodium taken from the advancing anterior margin, stained for cytoskeletal proteins. Note the dense, branching structure of each network and that they articulate onto each nucleus (nucleic acids stained blue). (a) Stained for actin with antibody ACTN(05) C4 (red). (b) Stained for tubulin with antibody KMX-1 (green). (c) Merge of figs. a + b with extended focus.

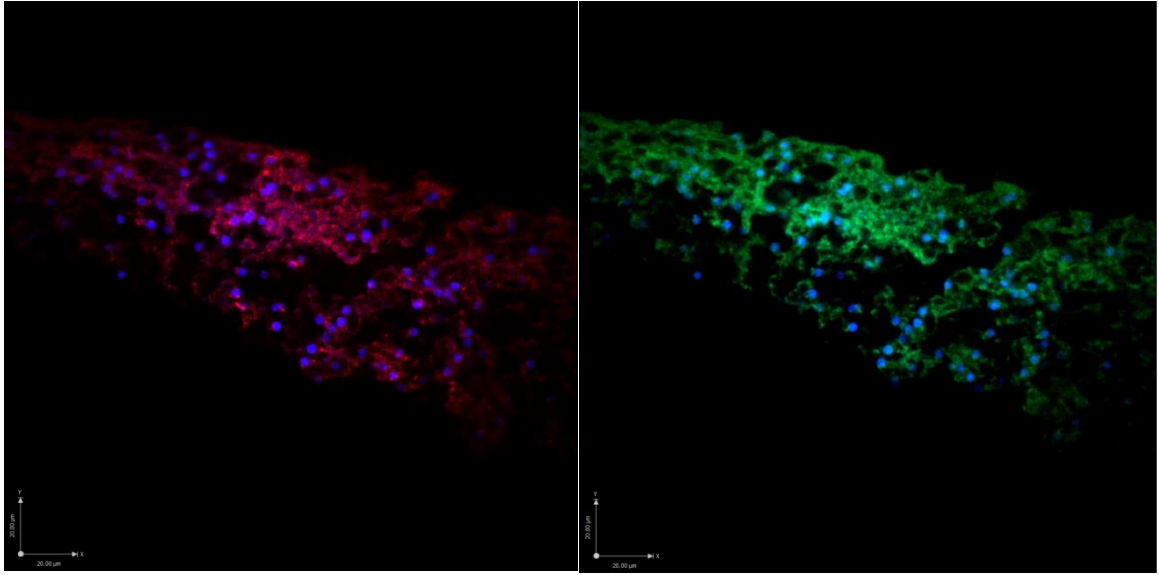

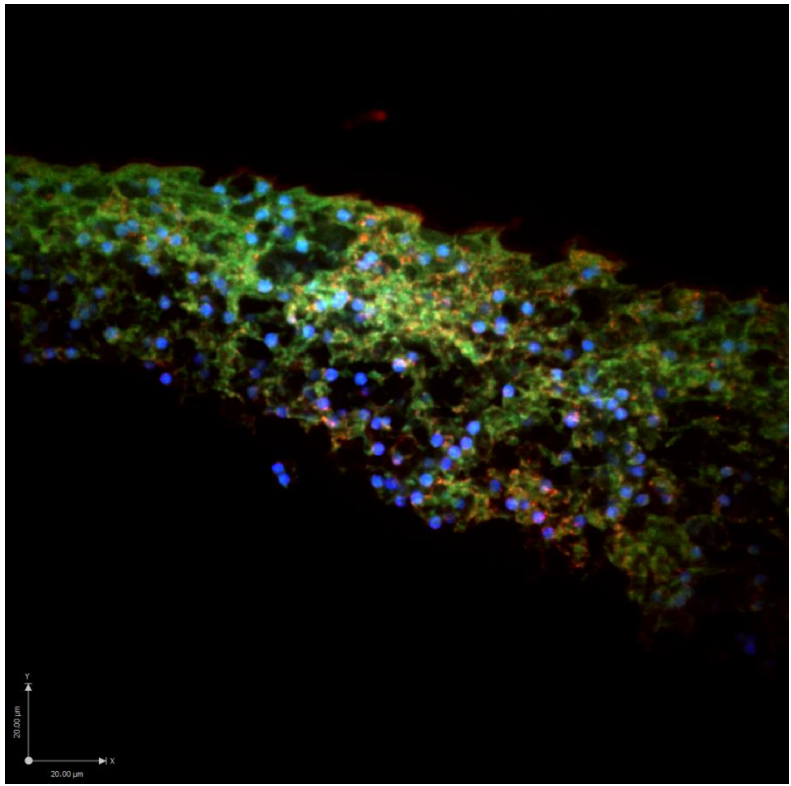

***Figure 4***

Higher magnification confocal micrographs of plasmodial sections stained for cytoskeletal proteins (same staining characteristics as Fig. 3). (a) Actin network. (b) Tubulin network surrounding two isolated nuclei.

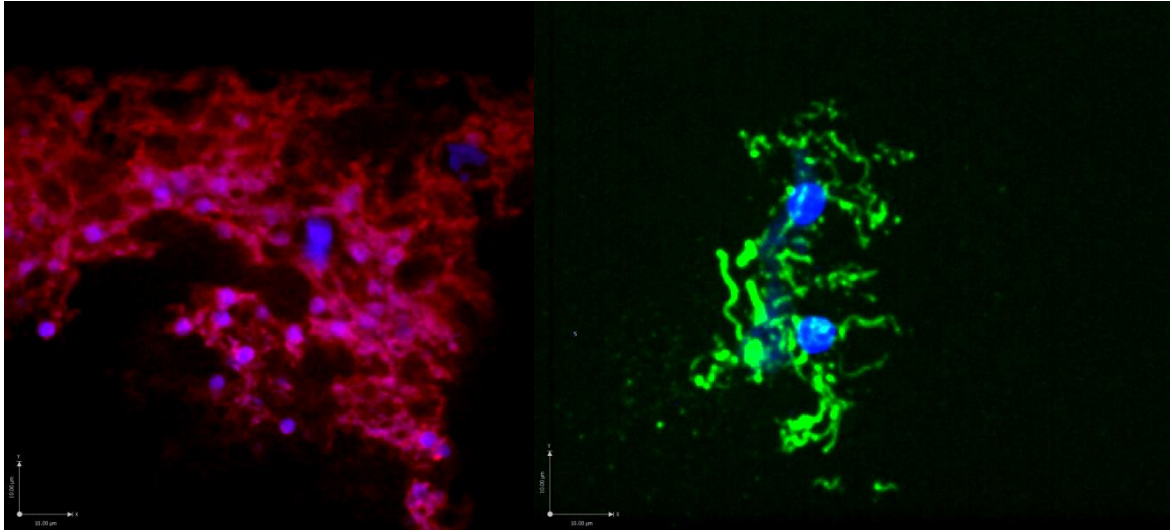

*Figure 5*

Proximity graphs as a potential formalisation of the communication and actuation network. The graphs are derived from experimental results Fig. 4a, assuming that nuclei are nodes. (a) Gabriel graph. (b) Relative neighbourhood graph. (c) Minimum spanning tree.

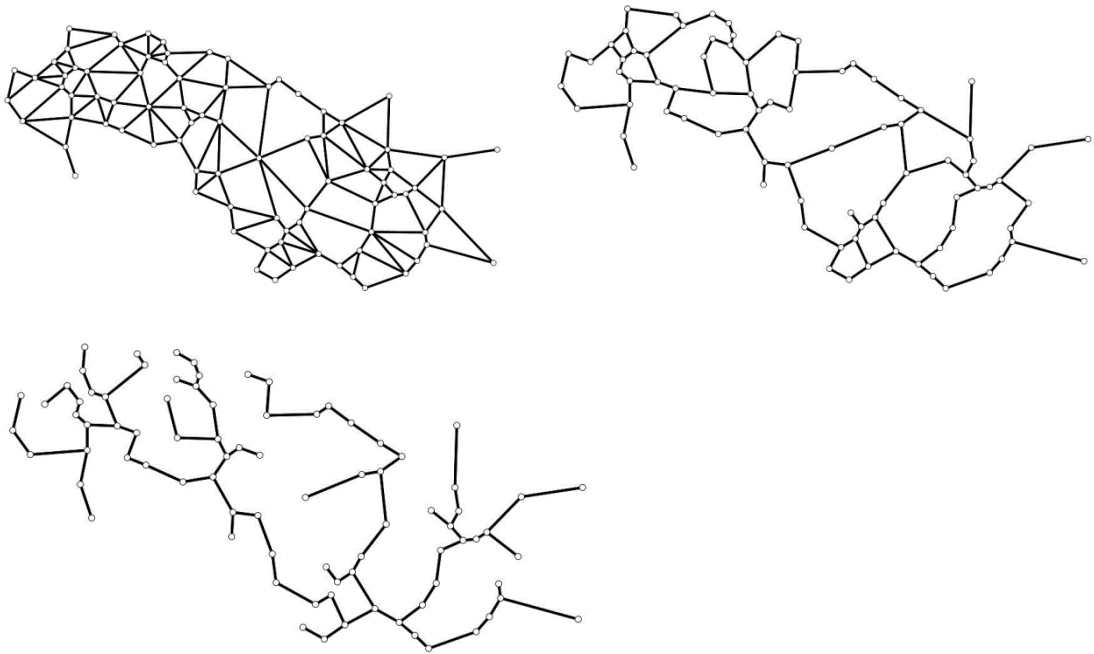

*Figure 6*

Schematic diagram of an f-actin strand where individual circles represent individual g-actin monomers. Cellular automaton rules are displayed for molecule $x_i$ (see text for rule description).

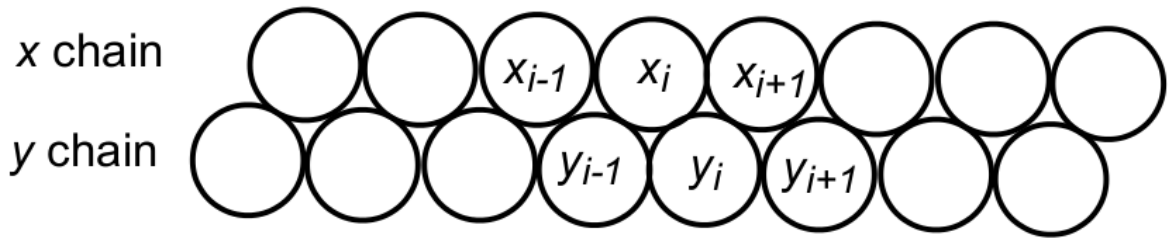

*Figure 7*

Actin automata evolving by rule $C_1()$. Original configuration is a randomly excited nodes, where every node takes state $+$ or $-$ with probability 0.25. Only space-time configurations of chain $x$ are shown. Time goes down. Excited nodes are shown by black pixels and resting nodes are white (no refractory nodes are present).

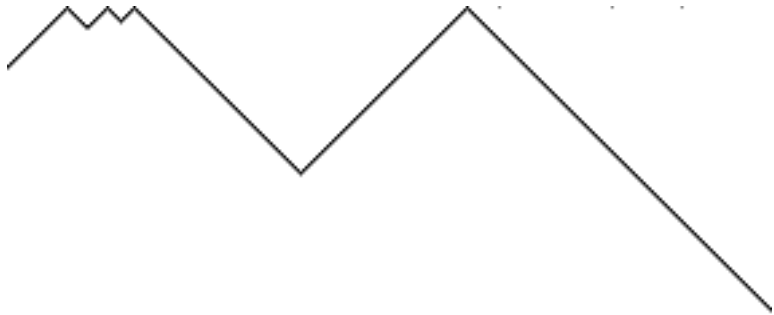

*Figure 8*

Actin automata evolving by rule $C_2()$. Original configuration is a randomly excited nodes, where every node takes state $+$ or $-$ with probability 0.25. Only space-time configurations of chain $x$ are shown. Time goes down. Excited nodes are shown by black pixels, refractory nodes by grey pixels and resting nodes are white.

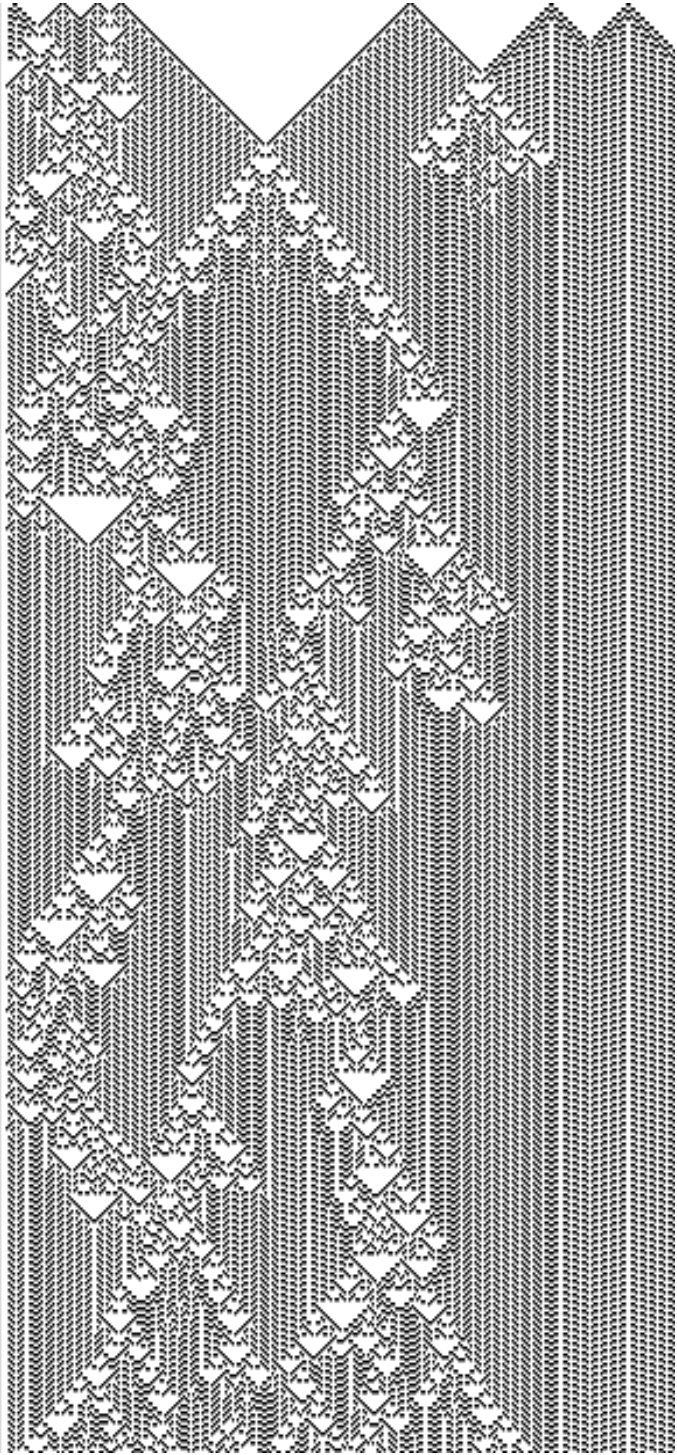

***Figure 9***

Actin automata evolving by rule $C_3()$. Original configuration is a randomly excited nodes, where every node takes state + or − with probability 0.25. (a) Space-time configurations of chain $x$, (b) Space-time configurations of chain $y$. Time goes down. Excited nodes are shown by black pixels, refractory nodes by grey pixels and resting nodes are white.

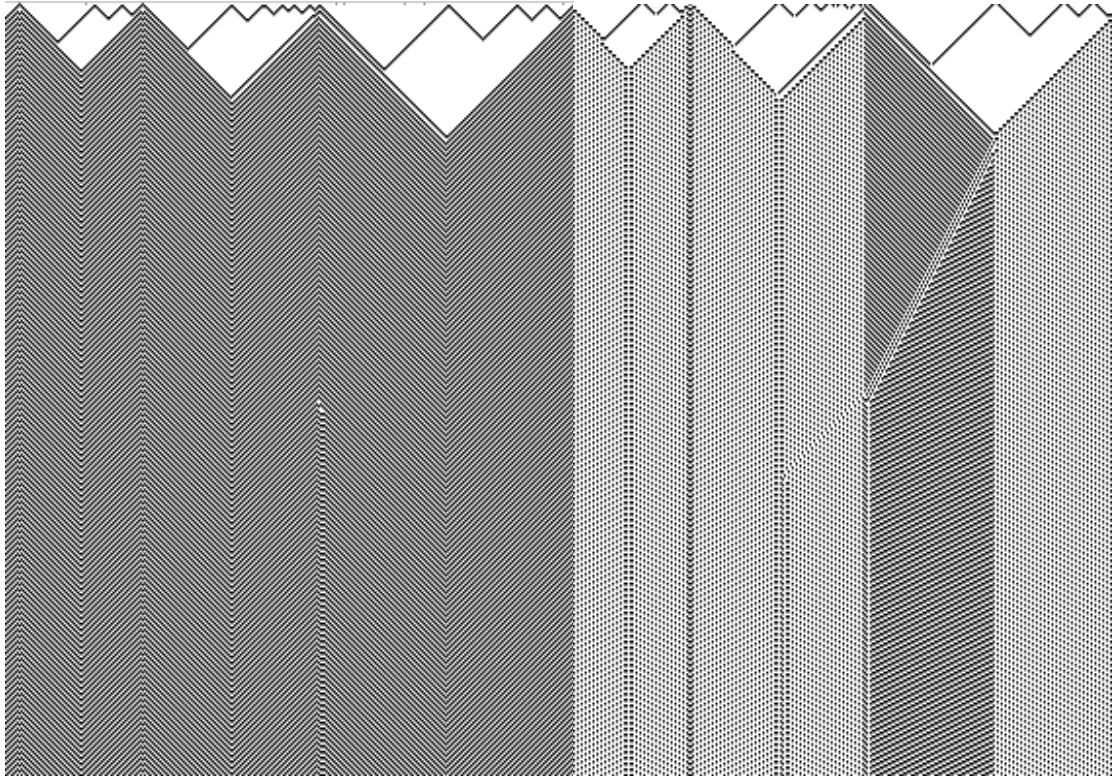

*Figure 10*

How a multi-agent model of slime mould outsources computation to the environment. (a) A small population (particle positions shown) of virtual plasmodium is inoculated on lowest node (bottom) and grows towards first node, engulfing it, and reducing chemoattractant projection, (b-d) Model population grows to nearest sources of chemoattractant completing construction of the spanning tree, (e-h) Visualisation of the changing chemoattractant gradient as the population engulfs and suppresses nutrient diffusion.

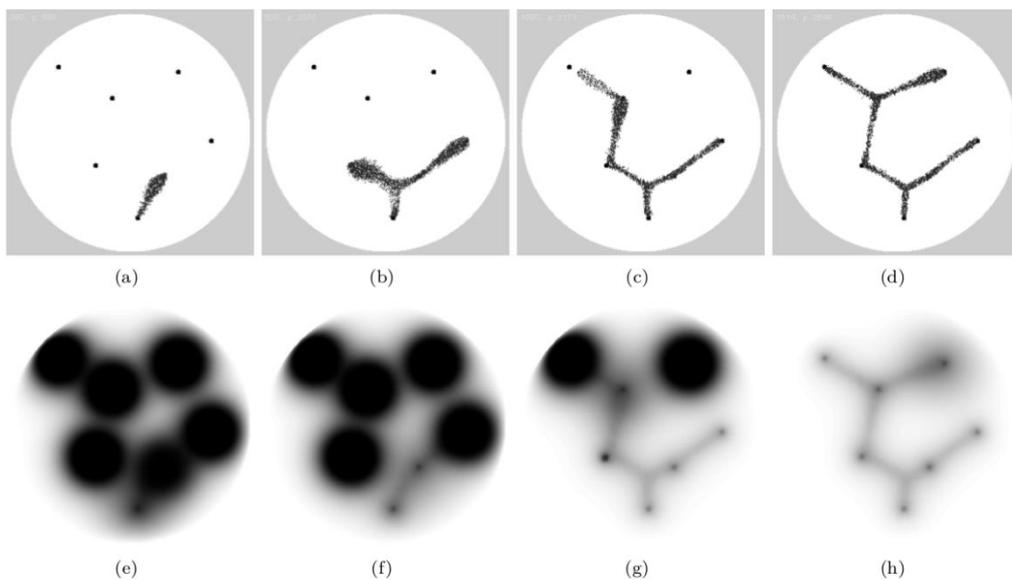